\documentclass[aps,prl,twocolumn,twoside,showpacs]{revtex4}
\usepackage{amsfonts}
\begin{document}

\title{
The Dirac field and the possible origin of gravity. }
\author{ A. Makhlin}
\email[]{amakhlin@comcast.net}
\affiliation{Rapid Research Co., Southfield, MI USA}
\date{August 6, 2004}
\begin{abstract}
The spin connections of the Dirac field have three ingredients that
are connected with the Ricci rotations, the Maxwell field, and an
{\em axial field} which minimally interacts with the axial current.
I demonstrate that the  axial field provides an  effective mechanism
of auto-localization of  the Dirac field into compact objects. The
condition that these objects are stable (the energy-momentum is
self-adjoint) leads to Einstein's field equations. The Dirac
field with its spin connection seem to be a natural material carrier
of the space-time continuum in which compact objects are moving
along geodesic lines. The long-distance effect of the axial
field is indistinguishable from Newton's gravity, which reveals
the microscopic nature of gravity and the origin of the gravitational
mass.
\end{abstract}
\maketitle

{\bf 1.} In this letter, I study the conditions under which the Dirac
field can form compact objects, i.e. particles that move along
world lines. The first key element of this study is the notion of
parallel transport of the Dirac field. All physical properties of
the spinor field are encoded in its tensor observables, some of which
are considered as internal. They should be transported with the
particle and cannot change if the particle is moving in free space.
The second key element is the requirement that the observables must
be represented by self-adjoint operators.

It is useful to note that the absolute differential, $DV_a \equiv
D_cV_a d s^c$, of a vector $V_a$ is the principal linear part of the
vector increment with respect to its change in the course of a
parallel transport along the same infinitesimal path; parallel
transport just means that $DV_a=0$.  The absolute differential of the
Dirac field is needed for exactly the same reason. A stable spinor
object does not change when it {\em is} parallel-transported along
its world line.

When dealing with spinors, one has to use the tetrad formalism
\cite{Eisenhart}, and the spinors should always be treated as
coordinate scalars. The Dirac matrices are pure number constructs
(I use them in the original form that was suggested by Dirac
\cite{Dirac1} ). The affine connections for vectors  can be derived
in a relatively simple way because the rotation of a vector at a given
point follows the rotation of the  local coordinate axes. There is no
similar rule for spinors since their components are not tensor
variables. Addressing this problem, I closely follow Fock's
method \cite{Fock1} and require the probability current, $j_a
=\psi^+ \alpha_a \psi =  \psi^+ (1,\rho_3\sigma_i)\psi$, to
be a Lorentz vector. This vector must be transformed as $j_a(x) \to
\Lambda_{a}^{~b}(x)j_b(x)$ under a local Lorentz rotation, and the
variation of its tetrad components under a parallel displacement
$ds^a$ should be $\delta j_a=\omega_{acb}j^c ds^b~$, where
$\omega_{bca}=-\omega_{cba}~$ are the Ricci rotation coefficients
fixed by the fact that the tetrad vectors
are covariantly constant, i.e.,
$D_\mu e^b_\nu =\nabla_\mu e^b_\nu-e^a_\mu \omega^b_{~ca}e^c_\nu=0$,
and $\nabla_\lambda e_\mu= \partial_\lambda e_\mu
-\Gamma^\nu_{\mu\lambda} e_\nu$.

Let the matrix $\Gamma_a$ (the spin connection) define the change of the
spinor components in the course of the same infinitesimal displacement,
$\delta \psi= \Gamma_a \psi ds^a$, $~\delta \psi^+=\psi^+ \Gamma^+_a ds^a$.
This gives yet another expression for $\delta j_a$,
$\delta j_a= \psi^+ (\Gamma^+_b \alpha_a+\alpha_a\Gamma_b)\psi ds^b$.
The two forms of $\delta j_a$ must be the same. Hence,
the equation that defines $\Gamma_a$ is
\begin{eqnarray}
\Gamma^+_b \alpha_a+\alpha_a\Gamma_b= \omega_{acb} \alpha^c ~,
\label{eq:E1}\end{eqnarray}
and its most general solution is,
\begin{eqnarray}
\Gamma_b(x)=-ieA_b(x)-ig\rho_3 \aleph_b(x)+ \Omega_b(x)~,
\label{eq:E2}\end{eqnarray}
where $A_a(x)$ and $\aleph_a(x)$ are real vector fields, and
$\Omega_b(x)=(1/4)\omega_{cdb}(x)\rho_1\alpha^c\rho_1\alpha^d$
is the geometric part.  The covariant derivative of a spinor now reads as
$D_a\psi=(\partial_a-\Gamma_a)\psi$ and
$D_\mu\psi=e^a_\mu D_a\psi=(\partial_\mu-\Gamma_\mu)\psi,~$
in the tetrad and coordinate basis, respectively.

The absolute differential of any observable spinor form $O$ which is
defined by an operator ${\cal O}$ is
\begin{eqnarray}
DO(x) \equiv  D [\psi^+ {\cal O} \psi]=
~\psi^+ ({\cal O}\overrightarrow {D_a} +
\overleftarrow{D^+_a}{\cal O}) \psi~ds^a~.
\label{eq:E3}\end{eqnarray}
I extensively use this rule throughout this work because it
{\em defines} a stable and, ultimately, compact spinor object.
The parallel transport of $O$ just means that $DO=0$.

If $~{\cal J}_a=\psi^+ \rho_3\alpha_a \psi~$  is the axial current,
and $~{\cal S}=\psi^+ \rho_1 \psi$ and $~{\cal P}=\psi^+ \rho_2 \psi$
are two Lorentz scalars then
\begin{eqnarray}
D_\mu j^\nu=\nabla_\mu j^\nu~,~~~~
D_\mu {\cal J}^\nu=\nabla_\mu {\cal J}^\nu~, \nonumber \\
D_\mu {\cal S}=\nabla_\mu {\cal S}+2g\aleph_\mu {\cal P},~
D_\mu {\cal P}=\nabla_\mu {\cal P}-
2g\aleph_\mu {\cal S}~.
\label{eq:E4}\end{eqnarray}
The first of these equations just duplicates the input for
Eq.~(\ref{eq:E1}). The parallel transport of vector
currents that are built with the aid of the diagonal Dirac matrices
$\sigma_i$ and $\rho_3$ is not affected by the axial field,
while for the Lorentz scalars, the left and right spinors are
mixed by either $\rho_1$ or $\rho_2$, which makes their covariant
derivative dependent on $\aleph_\mu$.

{\bf 2.} Taking the linear relation, $u_\mu P^\mu=m$, as a classical
prototype for the equation of motion, we can write the following
version of the Dirac equation ,
\begin{eqnarray}
\alpha^a(\partial_a \psi -\Gamma_a \psi)+
im\rho_1 \psi=0~.
\label{eq:E5}\end{eqnarray}
This equation includes an additional {\em axial} vector field
$\aleph_a(x)$, which is minimally coupled to the axial current and
acts differently on different spinor components.  The differential
operator of the Dirac  equation is Hermitian (symmetric), which is
confirmed by the conservation of the probability current $j^\mu$.
However, since in the most general form of the spin connection the
field $\aleph$ can be singular, it is not  necessarily a self-adjoint
operator. The Lorentz invariant condition that $i D_\mu$ is a
self-adjoint operator is as follows,
\begin{eqnarray}
i \int \partial_\sigma
(\sqrt{-{\rm g}}\psi^+ [\alpha^\sigma \overrightarrow{D_\mu} +
 \overleftarrow{D^+_\mu} \alpha^\sigma ] \psi)
 d^3 {\vec x} dx^0 \nonumber\\
=i \int \partial_\sigma \nabla_\mu
(\sqrt{-{\rm g}}\psi^+\alpha^\sigma\psi) d^3 {\vec x} dx^0 =0.
\label{eq:E6}\end{eqnarray}

{\bf 3.} The following identities directly result from the equation of motion
(\ref{eq:E5}) and its conjugate,
\begin{eqnarray}
\nabla_\mu j^\mu=0, ~{\rm and}~\nabla_\mu {\cal J}^\mu =2m {\cal P}.
\label{eq:E7}\end{eqnarray}
The first of these identities is the conservation of
the time-like probability current. It provides the definition of
a scalar product in the space of the Dirac spinors as an integral over
the three-dimensional space-like surface. By virtue of the first
of Eqs.~(\ref{eq:E4}) it also means
that the probability density {\em is} parallel-transported along the
vector of probability current, $D_\mu j^\mu=0$. The second identity
tells us that the axial current cannot be conserved (indeed,
${\cal J}^2=-j^2<0$; this vector is space-like).

Let us now introduce a standard energy-momentum tensor which is
the flux of the 4-momentum density,
\begin{eqnarray}
T^{\sigma}_{~\mu}=i\psi^+ \alpha^\sigma \overrightarrow{D_\mu} \psi~.
\label{eq:E8}\end{eqnarray}
Its operator must also be self-adjoint
so that the solutions of the Dirac equation and its adjoint belong
to the same space (e.g., have the same spectra of energies). This
condition has to be parallel-transported with the compact object that
``owns'' this spectrum, and the covariant form of this requirement is
\begin{eqnarray}
D_\sigma\{\psi^+ [ \alpha^\sigma \overrightarrow{D_\mu} +
 \overleftarrow{D^+_\mu} \alpha^\sigma ] \psi\}=0,
\label{eq:E9}\end{eqnarray}
which can be identically transformed into
\begin{eqnarray}
 \partial_\sigma( \sqrt{-{\rm g}}
[ \psi^+ \alpha^\sigma \overrightarrow{D_\mu} \psi \!+\!
\psi^+ \overleftarrow{D^+_\mu} \alpha^\sigma  \psi])\!
+\! R_{\mu\sigma}\! \sqrt{-{\rm g}}j^\sigma \!=\! 0,~
\label{eq:E10}\end{eqnarray}
where $R_{ad}={\rm g}^{bc}R_{ab;cd}$ is the Ricci curvature tensor.
Now, we integrate (\ref{eq:E10}) over the space-time domain, and compare
the result with the condition (\ref{eq:E6}).
This comparison shows that the Dirac field has a self-adjoint
Hamiltonian only if it lives in  {\em free space}. Since
$j^\mu\neq 0$, we must have
\begin{eqnarray}
R_{\lambda\sigma}=0~.
\label{eq:E11}\end{eqnarray}
In general relativity theory, this is Einstein's equation in
``empty'' space.

It is straightforward to show (e.g., using the technique
of Ref.~\cite{Fock1}) that, by virtue of the
equations of motion, the following identity holds
\begin{eqnarray}
\nabla_\sigma T^{\sigma}_{~\mu}= -eF_{\sigma\mu}j^\sigma\!
-g{\cal U}_{\sigma\mu}{\cal J}^\sigma \! -2gm\aleph_\mu{\cal P}
+{iR_{\mu\sigma}\over 2}j^\sigma,~
\label{eq:E12}\end{eqnarray}
where the commutator of the covariant derivatives is expressed
in terms of two  field strength tensors,
$ F_{\mu\nu}=\partial_\mu A_\nu -\partial_\nu A_\mu$ and
${\cal U}_{\mu\nu}=\partial_\mu \aleph_\nu -\partial_\nu \aleph_\mu$,
and the Ricci curvature tensor, $R_{ab}$.
By virtue of (\ref{eq:E11}) the last term in (\ref{eq:E12})
is zero, which renders the energy-momentum of the Dirac field a
real function.

{\bf 4.} The equations for the fields $A_a$ and $\aleph_a$ that
comply with  the data (hydrogen spectra) and with the position of
these  fields in the spin connection are as follows:
\begin{eqnarray}
\nabla_\sigma F^{\sigma\mu}=ej^\mu~,
\label{eq:E13}\end{eqnarray}
so that $F_{\mu\nu}$ is a massless gradient-invariant Maxwell
field which has the {\em probability current} as its source.
For the axial field $\aleph_\mu$ a plausible choice
is a massive  neutral vector field,
\begin{eqnarray}
\nabla_\sigma {\cal U}^{\sigma\mu}+M^2
\aleph^\mu=g {\cal J}^\mu~,
\label{eq:E14}\end{eqnarray}
which sufficiently describes the parity nonconservation
phenomena in atoms \cite{Khriplovich}. The Lorentz forces in
Eq.~(\ref{eq:E12}), with the field tensors
$F_{\mu\nu}$ and ${\cal U}_{\mu\nu}$ in a familiar position,
prompt the same equations of motion, because these equations
allow one to present the  Lorentz force  as the divergence of
the energy-momentum tensor. We have
\begin{eqnarray}
ej^\sigma F_{\sigma\mu}=
\nabla_\lambda[F^{\lambda\nu}F_{\nu\mu}+
{1\over 4}\delta^\lambda_\mu F^{\rho\nu}F_{\rho\nu}]
=\nabla_\lambda \Theta^\lambda_{~~\mu}~.
\label{eq:E15}\end{eqnarray}
Using (\ref{eq:E14}) and (\ref{eq:E7}) one can reduce
the Lorentz force of the axial field in (\ref{eq:E12}) to
\begin{eqnarray}
g{\cal J}^\sigma {\cal U}_{\sigma\mu}
+g \aleph_\mu~(\nabla_\sigma {\cal J}^\sigma)=
\nabla_\lambda (\theta^\lambda_{~\mu}+t^\lambda_{~\mu})~,
\label{eq:E16}\end{eqnarray}
where
\begin{eqnarray}
\theta^\lambda_{~\mu}\!\!=\!{\cal U}^{\lambda\nu}{\cal U}_{\nu\mu}\!+\!
{\delta^\lambda_\mu\over 4} {\cal U}^{\rho\nu}\!
{\cal U}_{\rho\nu},~
t^\lambda_{~\mu}\!\!=\!M^2(\aleph^\lambda \aleph_\mu \! -\!
{\delta^\lambda_\mu\over 2}\aleph^\rho \aleph_\rho).~
\label{eq:E17}\end{eqnarray}
In both (\ref{eq:E13}) and (\ref{eq:E14}) we followed a standard
convention that the charge is a divergence of electric
field. Then the positive charge corresponds to the positive flux of
the electric field outside a surrounding surface. With this convention,
the energy components $\Theta^{00}$, $\theta^{00}$ and $t^{00}$
of tensors (\ref{eq:E15}) and (\ref{eq:E17}) come out {\em
positive} exclusively because the coupling constants in the electron
spin connection (\ref{eq:E2}) are chosen {\em negative}.
Assembling Eqs.~(\ref{eq:E12})-(\ref{eq:E17}), one finds that
\begin{eqnarray}
\nabla_\lambda ( T^\lambda_{~~\mu} +\Theta^\lambda_{~~\mu}
+\theta^\lambda_{~~\mu}+t^\lambda_{~~\mu})=0~,
\label{eq:E18}\end{eqnarray}
i.e., the total energy-momentum of the interacting fields
$\psi$, $A_\mu$, and $\aleph_\mu$ is conserved, which is an additional
indication that the system of equations of motion is self-consistent.

{\bf 5.} Special relativity is built on two premises, i.e.,
the light-like propagation of all fields that carry  signals and
the existence of inertial frames. The first one is readily implemented by
a spinor representation of the Lorentz group: the two-component
Weyl spinors just map the light cone since their current is light-like.
The second one is more difficult to implement because the solutions
of the relativistic wave equations are not easily localized
to the extent where they can serve as the observers
(rods and clocks) of special
relativity. At the same time, the data indicate that the truly
localized interactions are due to the spinor fields.

An example of a relativistic point-like
particle with ``built in'' polarization $\vec{\zeta}$
 \cite{Weinberg} is helpful. If
$\vec{\zeta}$ of the rest frame is a part of a 4-vector
which is orthogonal to the 4-velocity, $\zeta^\mu u_\mu=0$, then
both  vectors are converted into vector fields {\em by means of}
their parallel transport,
$$ Du^\mu =(\partial_\nu u^\mu
+\Gamma^\mu_{\lambda\nu}u^\lambda)u^\nu d\tau=0,$$
$$D\zeta^\mu =(\partial_\nu \zeta^\mu
+\Gamma^\mu_{\lambda\nu}\zeta^\lambda)u^\nu d\tau=0.$$
These are the
equations for the geodesic trajectory of a point-like particle and
for the parallel transport of its polarization along this trajectory,
respectively. The vector $\vec{\zeta}$ is a  prototype for the
{\em internal polarization} of the Dirac spinor fields which is
represented by  various bilinear forms, like $\cal S$, $\cal P$,
etc. As long as we wish
to treat the electron as a compact object with an internal
polarization (quantum numbers including spin, charge, etc.)
we have to impose the {\em requirement} of parallel transport,
\begin{eqnarray}
D{\cal S}=0~,~~~~D{\cal P}=0~,~~\ldots
\label{eq:E19}\end{eqnarray}
for these quantities. It is imposed  on the forms in
which the left- and right- spinors are mixed by the matrices $\rho_1$
or $\rho_2$. The Ansatz (\ref{eq:E19}) serves as an additional
condition on the spin connection $\rho_3\aleph_a(x)$ of the Dirac field, which
makes $\aleph_a$ compatible with the existence of a freely moving stable
spinor particle. With an {\em a priori} form (\ref{eq:E2}) of the spin
connection,  (\ref{eq:E19}) can be cast into the following coordinate form,
\begin{eqnarray}
D_\mu \!{\cal S}\!=\!\nabla_\mu {\cal S}\!+
\! 2g\aleph_\mu {\cal P}\!=\!0,~~
D_\mu \!{\cal P}\!=\!\nabla_\mu {\cal P}
\!-\!2g\aleph_\mu {\cal S}\!=\!0.~
\label{eq:E20}\end{eqnarray}
Now, instead of being an arbitrary external field that acts on a
Dirac spinor, the field $\aleph_\mu$ becomes a
functional of spinor forms. In order to find this functional note that from
Eqs.~(\ref{eq:E20}) it follows that
$~{\cal S}\partial_\mu{\cal S}+{\cal P}\partial_\mu{\cal P}= 0$.
Hence, the scalar function~ ${\cal R}^2={\cal S}^2 +{\cal P}^2$,
is the first integral of these two equations.  Since ${\cal S}$ and
${\cal P}$ are real functions, we may look for  solutions in the
forms of
${\cal S}={\cal R}\cos\Upsilon$, and ${\cal P}=-{\cal R}\sin\Upsilon$,
which yields the following equation,
\begin{eqnarray}
 2g \aleph_\mu  = - \partial_\mu \Upsilon~.
\label{eq:E21}\end{eqnarray}
 Because ${\cal R}^2=j^aj_a$ is the squared probability current, a
positive ${\cal R}$ is a natural measure of the {\em localized}
spinor matter.  {\em The potential function $\Upsilon(x)$
remains a distinct characteristic of the Dirac field
even in that part of space where ${\cal R}\to 0$.}
For perfectly stable spinor matter, the axial field $\aleph_\mu$
is completely defined by one
scalar function. Its curvature tensor vanishes, $~{\cal U}^{\sigma\mu}=0$.
Furthermore, if the conditions (\ref{eq:E19}) are exact, then
the Eqs.~(\ref{eq:E17}) become $~\theta^\lambda_{~\mu}=0$, and
\begin{eqnarray}
t_{\lambda\mu}=(M^2/ 8 g^2)
\big( 2~ \partial_\lambda\Upsilon \partial_\mu\Upsilon -
g_{\lambda\mu}\partial_\rho\Upsilon
\partial^\rho\Upsilon \big).~~
\label{eq:E22}\end{eqnarray}

The equation for the field $\Upsilon(x)$ can be found by computing
the covariant divergence of Eq.~(\ref{eq:E21}) and expressing
the divergence of the axial field via the pseudoscalar density,
$M^2\nabla_\mu \aleph^\mu=2gm{\cal P}=-2gm{\cal R}\sin\Upsilon $,
i.e.,
\begin{eqnarray}
\Box~ \Upsilon(x)-(4g^2m/M^2)~{\cal R}(x)\sin\Upsilon(x)=0~.
\label{eq:E23}\end{eqnarray}
Within a domain where ${\cal R}(x)$ is  constant, this is the
well-known sine-Gordon equation which has soliton-type solutions.
This equation is classical, the Planck constant has cancelled out;
therefore, the field $\Upsilon(x)$ does not vanish in the classical limit.
In the case of a variable source ${\cal P}$, the variations of $\Upsilon(x)$
propagate according to the d'Alembert equation. The Dirac equation itself
becomes
\begin{eqnarray}
\alpha^a \{\partial_a
+ieA_a+{i\over 2}~\rho_3\partial_a\Upsilon
 -\Omega_a\}\psi +im \rho_1\psi=0,
\label{eq:E24}\end{eqnarray}
and we have a {\em nonlinear system} of Eqs.(\ref{eq:E23})-(\ref{eq:E24})
in which the scale parameter $m$ can no longer be arbitrary.
Furthermore, the Dirac equation is singular. Indeed, let
Eq.(\ref{eq:E23}) has a spherically symmetric static solution.
If the exterior of the domain $r<r_{max}$ is an empty space where
${\cal R}^2=j_0^2=0$, then the function $\Upsilon$ is a solution of
the external problem for the Laplace equation  and when $r>r_{max}$ we have
\begin{eqnarray}
\Upsilon(r)= - {1\over r}~{4g^2 m\over M^2}~
\int_0^{r_{max}}\!\!\!\! {\cal P}(r)r^2~dr=-{{\cal Q}\over r},
~\big({\cal Q}\stackrel{>}{<} 0\big).~
\label{eq:E25}\end{eqnarray}
The potential, which is as singular as
$-\partial_r\Upsilon\propto -{\cal Q}/r^2$, in the
Dirac equation brings about the advent of the ``falling onto the centre''
phenomenon. This leads to a super-critical binding which initiates
a spectrum of (quasi)bound states where the Dirac field can
have negative energy.

{\bf 6.} Following the same logic, we have to require that a compact
spinor object localizes its energy and momentum along its world line
and to add  a new element,
\begin{eqnarray}
D_\sigma T^\sigma_\mu=0~,
\label{eq:E26}\end{eqnarray}
to the Ansatz (\ref{eq:E19}). The kinetic 4-momentum
is parallel-transported with the particle  when
it is moving in free space. Then, by definition,
\begin{eqnarray}
D_\sigma T^\sigma_{~\mu}\equiv i~
 (\psi^+ \overleftarrow{D^+_\sigma}\alpha^\sigma
 \overrightarrow{D_\mu} \psi +
\psi^+ \alpha^\sigma \overrightarrow{D_\mu}
\overrightarrow{D_\sigma} \psi) =0~,
\label{eq:E27}\end{eqnarray}
which can be identically transformed to
\begin{eqnarray}
\partial_\sigma(\sqrt{-{\rm g}}T^\sigma_{~\mu})=
i \sqrt{-{\rm g}}\psi^+ \alpha^\sigma[D_\mu D_\sigma -D_\sigma D_\mu]\psi,
\label{eq:E28}\end{eqnarray}
and compared with the identity (\ref{eq:E12}) that follows from the
equations of motion,
\begin{eqnarray}
(-{\rm g})^{-1/2} \partial_\sigma[\sqrt{-{\rm g}}T^{\sigma}_{~\mu}]
=\Gamma^\sigma_{\mu\nu}
T^\nu_{~\sigma}-2mg{\aleph}_\mu {\cal P}
~~~~~\nonumber\\
+i\psi^+ [D_\sigma D_\mu -D_\mu D_\sigma ]\psi.\nonumber
\end{eqnarray}
These two equations coincide if the force of inertia and the external
force from the axial field $\aleph$ are equal,
\begin{eqnarray}
 \Gamma^\sigma_{\mu\nu}T^\nu_{~\sigma} =
2mg  {\cal P}{\aleph}_\mu
=- m {\cal P}~\partial_\mu \Upsilon ~.
\label{eq:E29}\end{eqnarray}
While the first part (\ref{eq:E19}) of the Ansatz simplifies
the spin connection ${\aleph}_\mu$ to the gradient of a scalar
function, the second part (\ref{eq:E26}) specifies the affine
connection. The simplest form of Eq.~(\ref{eq:E29}) is
$T^{00}\partial_i {\rm g}_{00}=2m{\cal P}\partial_i\Upsilon $,
which immediately leads to
\begin{eqnarray}
{\rm g}_{00}=1+2(m{\cal P}/ T_{00})~\Upsilon \to
1+2\Upsilon_{grav}~,
\label{eq:E30}\end{eqnarray}
Thus, we have {\em derived} the key formulae (\ref{eq:E11}) and
(\ref{eq:E30}), which usually {\em are postulated} as the starting point
of general relativity. Therefore, all fundamental predictions of
general relativity are preserved. The field $\aleph_\mu$, is
indistinguishable from the gravitational field and is locally
equivalent to the field of inertial forces. Eq.~(\ref{eq:E29})
explicitly states  that it is possible to choose such a space-time
coordinates (metric) that a small body moves along the geodesic lines of
{\em this} metric.

{\bf 7.} A simple model with the axial potential (\ref{eq:E25})
and only one mass/energy parameter appears to be solvable exactly.
Large and small components of the {\em radially polarized}
and spherically symmetric modes of the Dirac field behave as $r\to 0$
like $F\sim\cos({\cal Q}/r)$ and $G\sim\sin({\cal Q}/r)$,
so that the probability density $R=r^2j^0=F^2+G^2~$ remains
nearly constant within the range of sharp localization.
These normalizeable bound states have a continuous energy spectrum
when $-\infty<E{\cal Q}<-3/8$. For the Dirac field with this simplest
geometry we have $d(F^2+G^2)/dr=4mFG=2m{\cal P}$,
which provides a simple formula
for the gravitational mass in Eqs.~(\ref{eq:E25}), (\ref{eq:E29})
and (\ref{eq:E30}), i.e., $$2m\int{\cal P}dV = R(0),$$
and allows one to conclude that, at the microscopic level,
the gravitational mass is proportional to the {\em peak amplitude of
the localization} of the Dirac field at the gravitating centers.
The smallest particles are the heaviest, and on a purely
dimensional ground we have $ R(0)\sim |E|$. This is evidence
that, even being of a different physical origin, the
inertial ($\sim \int T^{00}dV\sim E$) and the gravitational
($\sim m\int{\cal P}dV\sim R(0)$)
 masses of stable particles are the same up to a
possible factor which can be absorbed into ``Newton's gravitational
constant'' $G_N \sim (g^2/M^2)$; this suffices to guarantee the
universality of free fall. Only an extensive study of
Eqs.~(\ref{eq:E23}) and (\ref{eq:E24}) can clarify whether
this factor is universal. It looks like the mass, which corresponds
to this gravitational constant, is smaller than the formal
``dimensional'' Planck mass, $M_{Planck}^2=(\hbar c/G_N)$, by a factor
of $g$ related to the electro-weak interactions. We suggest that the
electro-weak and gravitational constants uniquely determine the mass
$M$ of the axial field.

{\bf 8.}  Equations (\ref{eq:E23}), (\ref{eq:E24}) and
(\ref{eq:E11}), (\ref{eq:E29}) clearly
support an image of particles as moving singularities
which {\em are shared by the  Dirac and gravitational fields}.
These equations guarantee that the theory is protected from {\em
mathematical singularities} which inevitably appear when the
nonlinear Einstein's equations (\ref{eq:E11}) are solved
independently \cite{EI}. A striking agreement between the character
of the motion of singular domains of the Einstein and the Dirac fields seems
to be an indication that general relativity naturally requires
matter in the form of the Dirac field. No other fields can
provide the degree of localization, which is necessary for such an
agreement. The complementarity of these two fields indeed solves the
problem of motion as it was posed by Einstein \cite{EI}.
In agreement with Einstein's eventual
judgement, the field equation (\ref{eq:E11}) has no energy-momentum
tensor of matter in its r.h.s.  {\em The physical origin of the
macroscopic forces of gravity between any two bodies is a trend of the
global Dirac field to concentrate around the microscopic domains
where this field happens to be extremely localized.}
These forces tend to polarize the matter at the level of its spinor
organization and they well may contribute at various stages of
matter evolution. The picture of gravity as the effect of the axial field
explains gravity as a coherent effect that cannot be screened
by any bodies or fields. The ``falling onto a centre''
is universal and is guaranteed by a special position of the singular
Newton's force as a potential in the Dirac equation.   It
would be fair to say that the Dirac field is such a natural ``stuffing''
for Einstein's singularities, that the prevailing  feeling of the
incompleteness of the Einstein-Infeld theory fades away.
Of all possible solutions of the Einstein's equations, only
those that have the material partners among the compact solutions
of the Dirac equation with the axial field in
spin connection, are physically meaningful.

\end{document}